# The Useful Side of Motion: Using Head Motion Parameters to Correct for Respiratory Confounds in BOLD fMRI[*]

Abdoljalil Addeh [1-3,5], G. Bruce Pike,[3-5] M. Ethan MacDonald [1-3,5]

[1] Department of Biomedical Engineering, Schulich School of Engineering, University of Calgary, Canada
[2] Department of Electrical & Software Engineering, Schulich School of Engineering, University of Calgary, Canada
[3] Department of Radiology, Cumming School of Medicine, University of Calgary, Canada
[4] Department of Clinical Neurosciences, Cumming School of Medicine, University of Calgary, Canada
[5] Hotchkiss Brain Institute, Cumming School of Medicine, University of Calgary, Canada

**Target audience:** MR motion correction specialists and neuroimaging researchers, those interested in motion and pseudo-motion of fMRI.

**Purpose:** Acquiring accurate external respiratory data during functional Magnetic Resonance Imaging (fMRI) is challenging, prompting the exploration of machine learning methods to estimate respiratory variation (RV) from fMRI data (Addeh et al., 2023). Respiration induces head motion, including real and pseudo motion, which likely provides useful information about respiratory events (Power et al., 2019). Recommended notch filters mitigate respiratory-induced motion artifacts, suggesting that a bandpass filter at the respiratory frequency band isolates respiratory-induced head motion (Fair et al., 2020; Kaplan et al., 2022).

This study seeks to enhance the accuracy of RV estimation from resting-state BOLD-fMRI data by integrating estimated head motion parameters. Specifically, we aim to determine the impact of incorporating raw versus bandpass-filtered head motion parameters on RV reconstruction accuracy using one-dimensional convolutional neural networks (1D-CNNs). This approach addresses the limitations of traditional filtering techniques and leverages the potential of head motion data to provide a more robust estimation of respiratory-induced variations.

**Methods:** We utilized 1D-CNNs to reconstruct the RV waveform from BOLD signals and head motion parameters. A total of 900 resting-state fMRI data and corresponding respiratory records from the Human Connectome Project in Young Adults (HCP-YA) dataset were employed for training and testing the proposed method (Van Essen et al., 2013). The fMRI images were corrected for geometric distortions caused by B0 inhomogeneity using the TOPUP tool in FSL; corrected for head motion by the MCFLIRT tool in FSL, with head motion parameters estimated directly from the raw fMRI data; and registered to 2 mm MNI space using Advanced Normalization Tools (ANTs). For computational efficiency and to boost the signal-to-noise ratio of the input signals, the mean BOLD signal time series from 90 functional regions of interest (ROI) is utilized as the primary model input (Shirer et al., 2012). The 90 ROIs encompassing a vast portion of the cortical and subcortical gray matter.

The model input is formulated using a sliding time-window approach. Specifically, fMRI-ROI signals and estimated head motion parameters are fragmented into brief, overlapping time-windows of 65 TRs (TR

---

= 0.72 s). For each time window, the model outputs a single time-point estimate of the RV signal at the first point, middle point, and end point of the window. The RV is defined as the standard deviation of the respiratory waveform within a six-second sliding window centered at each time point (Chang et al., 2009).

**Results:** Three experimental setups were tested: (1) using BOLD signals alone, (2) using BOLD signals combined with raw head motion parameters, and (3) using BOLD signals combine with bandpass-filtered head motion parameters as inputs to the CNN. Figure 1 shows an example of respiratory signal, raw head motion, and bandpass filtered head motion parameters. There was no significant difference between using only BOLD signals and combining BOLD signals with bandpass-filtered head motion parameters (p-value >0.01). However, results indicated that incorporating raw head motion parameters significantly improved RV reconstruction accuracy. Compared to using only BOLD-fMRI data, the inclusion of raw head motion parameters improved mean absolute error by 14%, mean square error by 24%, correlation by 14%, and dynamic time warping by 12%.

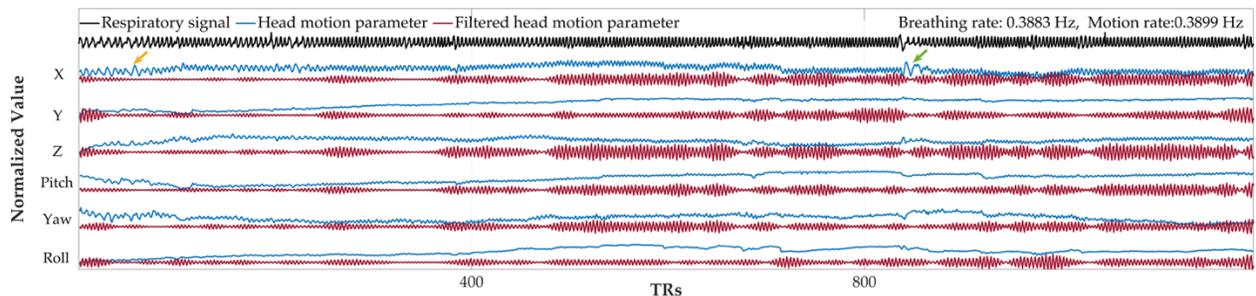

**Figure 1.** Example of respiratory signal, raw head motion parameters, and bandpass-filtered head motion parameters. The respiratory rate closely aligns with the frequency of the head motion parameters over the duration of the scan, illustrating the efficacy of head motion parameters in capturing respiratory information. Variability in respiratory rate and depth is observed throughout the fMRI scan, indicating occurrences of events such as 'deep breath' (green arrow) or 'shallow breathing' (orange arrow). This highlights the dynamic nature of respiration and underscores the importance of considering these variations in notch/bandpass filter design.

**Discussion:** Breathing rates can fluctuate considerably during an fMRI scan, making it challenging to capture respiratory components accurately with fixed notch or bandpass filters. These filters only work when their central frequency matches the actual breathing rate, leading to mismatches during events like deep breaths or slow breathing, which affect the BOLD signal differently. Deep breaths reduce the BOLD signal, while slow breathing increases it, causing inconsistencies that can mislead machine learning models. We recommend using raw head motion parameters, as they contain the full spectrum of motion-related information. Deep learning architectures, capable of sophisticated nonlinear filtering, are better suited to leverage patterns within unprocessed data for RV reconstruction.

**Conclusion:** Our findings reveal that the inclusion of raw head motion parameters, but not bandpass-filtered ones, substantially improves RV reconstruction accuracy. This result highlights the limitations of fixed-frequency bandpass filtering in capturing respiratory-related motion. This methodological improvement has profound implications, particularly in scenarios where standard respiratory monitoring is challenging or impractical, such as with pediatric, elderly, or diseased populations. By leveraging these advancements, researchers can achieve more accurate physiological measurements, enhancing the reliability and applicability of fMRI studies in diverse clinical and research settings.